\documentstyle[11pt,aaspp4]{article}
\received{1997 August 19}
\accepted{1997 October 28}
\lefthead{Rothschild et al.}
\righthead{HEXTE In-flight Performance}

\begin{document}

\newcommand{\ltsim}{\lower.5ex\hbox{$\; \buildrel < \over \sim \;$}}

\title{In-flight Performance of The High Energy X-Ray Timing Experiment on the Rossi X-ray Timing Explorer}

\author{R. E. Rothschild, P. R. Blanco, D. E. Gruber, W. A. Heindl, D.
R. MacDonald, D. C. Marsden,  M. R. Pelling, L. R. Wayne\\}

\affil{Center for Astrophysics and Space Sciences 0424, \\
University of California at San Diego, La Jolla, CA 92093$-$0424\\}

\and

\author{P. L. Hink}

\affil{Dept. of Physics, \\
Washington University, St. Louis, MO 63130}

\begin{abstract}

The High Energy X-ray Timing Experiment (HEXTE) is one of three
scientific instruments aboard the {\it Rossi X-ray Timing Explorer
(RXTE)}, which was launched on December 30, 1995.  {\it RXTE} performs
timing and spectral studies of bright x-ray sources to determine the
physical parameters of these systems.  The HEXTE consists of two
independent clusters of detectors, each cluster containing four
NaI(Tl)/CsI(Na) phoswich scintillation counters sharing a common
1$^{\circ}$ FWHM field of view. The field of view of each cluster is
switched on- and off-source to provide near real-time background
measurements. The net open area of the eight detectors is 1600 cm$^2$,
and each detector covers the energy range 15$-$250 keV with an average
energy resolution of 15.4\% at 60 keV.  The in-flight performance of
the HEXTE is described, the light curve and spectrum of the Crab
Nebula/Pulsar is given, and the 15$-$240 keV spectrum of the weak
source, active galaxy MCG 8-11-11 is presented to demonstrate the weak
source spectral capabilities of HEXTE.

\end{abstract}


\section{Introduction}

\begin{figure}
\plotone{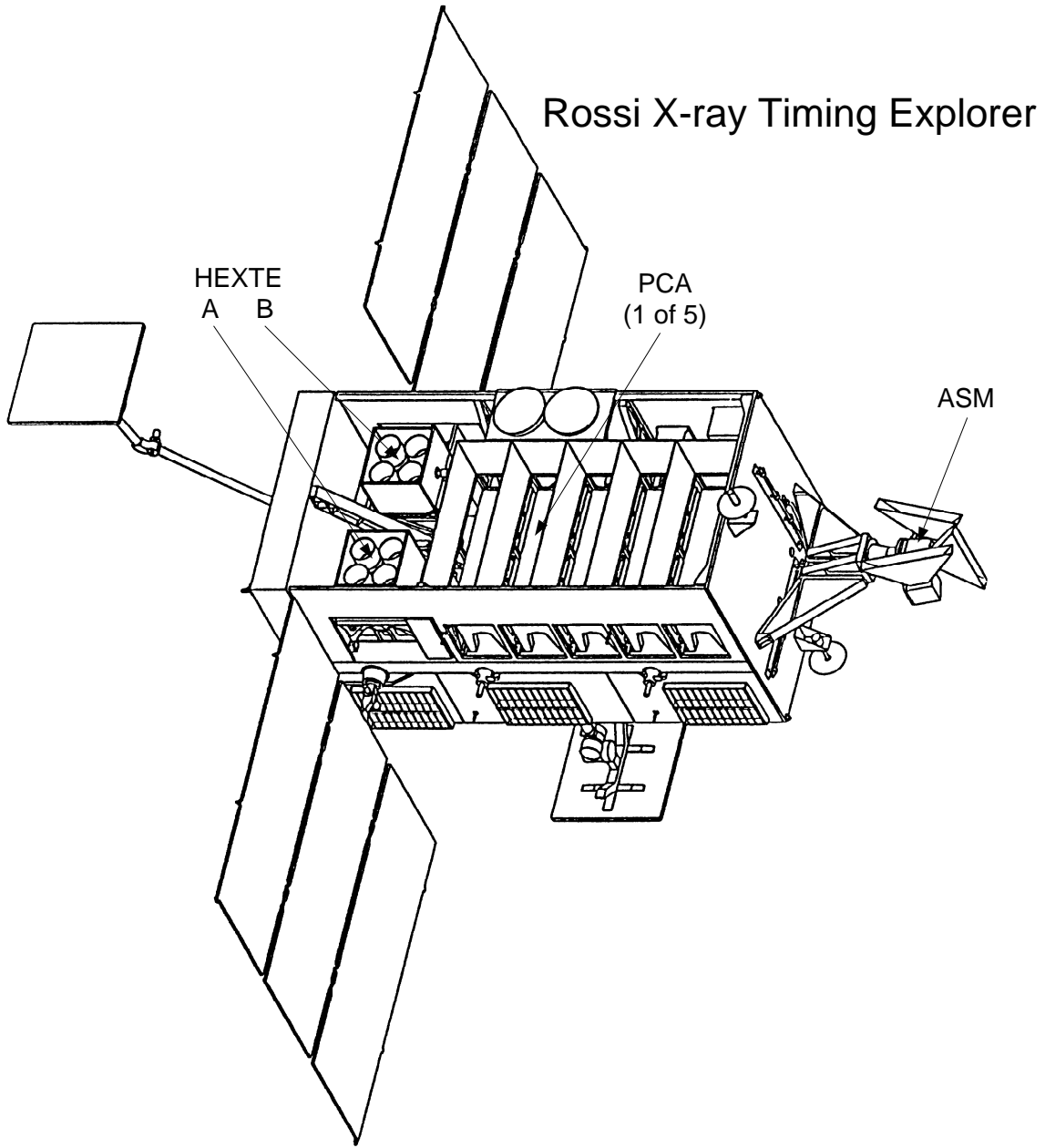}
\caption{The {\it RXTE} spacecraft viewed from above to
reveal the scientific instruments. The five PCA proportional counters
and the two HEXTE clusters are shown in relation to the rest of the
spacecraft components, such as the two high gain antennae for
communications with the ground via the Tracking and Data Relay
Satellite System, the two solar-power arrays that can rotate to face
the sun, and the ASM on the end with clearance to view the sky.\label{fig.rxte}}
\end{figure}

The High Energy X-ray Timing Experiment (HEXTE) is one of three
scientific instruments aboard the {\it Rossi X-ray Timing Explorer
(RXTE)} (see Fig. \ref{fig.rxte}). The other instruments are the
Proportional Counter Array (PCA, Jahoda et al.\markcite{jahoda} 1996)
and the All-Sky Monitor (ASM, Levine et al.\markcite{levine} 1996).
The PCA and HEXTE are co-aligned to provide 2$-$250 keV observations of
individual sources, and the ASM independently monitors the $\sim$100
brightest 2$-$12 keV sources in the sky (Swank et al. \markcite{swank}
1995).  The HEXTE is the latest in a long term program of hard
x-ray instrumentation development for observational high energy
astrophysics that has included the UCSD Cosmic X-ray Telescope on {\it
OSO-7} and the A-4 Hard X-ray and Low Energy Gamma-Ray Experiment on
{\it HEAO-1}. The successful launch of {\it RXTE} on a Delta II
expendable launch vehicle occurred at 13:47 UT on December 30, 1995.
The first month of operations was devoted to In-Orbit Check-out (IOC), the
results of which are described here.

\section{Instrument Description}

The HEXTE consists of two independent clusters of detectors. Each
cluster contains four NaI(Tl)/CsI(Na) phoswich scintillation counters
collimated by a lead honeycomb.  All eight collimators are co-aligned
on-source to give both clusters a 1$^{\circ}$ FWHM field of view.  The
net open area of the eight detectors is $\sim$1600 cm$^2$, and each
detector covers the energy range 15$-$250 keV with an average energy
resolution of 15.4\% FWHM at 60 keV.  The cluster organization for the
HEXTE instrument permits effective immunity to systematic background
variations through the use of continuous gain control, chopping of the
source signal, and anticoincidence shielding of charged particle
events. The system requires about 45 watts, exclusive of heater power,
weighs about 400 kg, and utilizes 5 kb/s of telemetry on the average.
Table 1 summarizes the HEXTE instrument characteristics.

The energy, time of arrival, and pulse shape associated with each
detected photon are measured and digitized. For sources weaker than the
Crab Nebula, all of this information may be telemetered to the ground.
The maximum anticipated event rate for the brightest sources, however,
does not allow all of this information to be transmitted, as each HEXTE
cluster's data rate is limited to 23,000 bits/s. Thus, the on-board
data processor permits, on command, the selection of any desired subset
of each event's data to be telemetered.  For very bright sources the
processor has two binned analysis modes to permit considerable
flexibility in data formatting and compression in order to achieve the
timing and energy resolution requirements for source observations
consistent with efficient data recovery.

The HEXTE instrument is sensitive to x-ray fluxes on timescales from
milliseconds to days, limited only by the telemetry and the source
intensity. HEXTE is capable of measuring a typical, weak ($\sim$1
milliCrab) x-ray source to 100 keV or greater in 10$^5$ live-seconds,
i.e., it has a 3$\sigma$ sensitivity of 10$^{-6}$ photons cm$^{-2}$ s$^{-1}$
keV$^{-1}$ in a 20 keV band at 100 keV.

\subsection{X-ray Detectors}

\begin{figure}[t]
\plotone{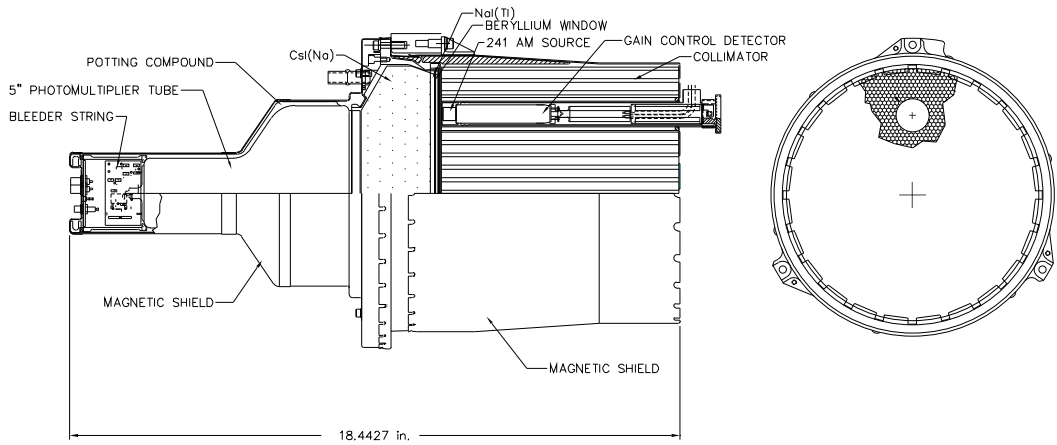}
\caption{A HEXTE detector module, consisting of photomultiplier
tube, scintillation crystals, and collimator assemblies. The collimator
assembly also contains the gain control photomultiplier and $^{241}$Am
doped plastic scintillator assembly. The upper half of the side view on
the left is a cut-away revealing the inner elements of the module,
while the lower half shows the magnetic shield housing that encases the
entire module. The front view on the right gives the placement of the
automatic gain control assembly in the collimator and shows a fraction of
the collimator honeycomb. \label{fig.detmodule}}
\end{figure}

The primary HEXTE detection elements are eight 18.3 cm diameter by
0.3 cm thick NaI(Tl) scintillation crystals each of which is
optically coupled to a single 5 inch photomultiplier tube through an
intervening 5.69 cm thick CsI(Na) shield crystal and 0.64 cm thick
quartz window, as shown in Figure \ref{fig.detmodule}. The rear of the
CsI(Na) crystal is tapered to match the photomultiplier tube's
photocathode diameter (typically 11.4 cm). The CsI(Na) crystal provides
uniform collection of primary NaI(Tl) crystal scintillation light and
active anticoincidence shielding, both from x-ray events not
originating in the look direction, and from events with partial energy
loss in the NaI(Tl).

The scintillation pulses generated within the two crystal media exhibit
differing characteristic decay times, 0.25 $\mu$s in NaI(Tl)
and 0.63 $\mu$s in CsI(Na). Signals from the photomultiplier tube are
pulse shape analyzed to distinguish pure NaI(Tl) energy losses from
events containing some proportion of the slower component indicating an
energy loss in the CsI(Na) shield crystal. Rejected events can be
either charged particles that trigger both crystals or x-rays that
leave partial energy in the NaI(Tl) and CsI(Na) crystals.

The crystals are contained in an opaque, hermetically sealed housing to
prevent degradation of the NaI(Tl) by water vapor before launch and to
shield the photomultiplier tube from stray light. The housing
incorporates a 0.051 cm thick beryllium x-ray entrance window to
provide a light seal with minimal low energy absorption of incident
x-rays. The crystals are wrapped in teflon sheet, 0.025 cm thick, and
are highly polished to provide uniform light collection by the
photomultiplier tube, thereby maximizing energy resolution.

The photomultiplier tube and its attendant high voltage divider network
(bleeder string), coupling elements, and connectors are encased in
silicone elastomer, and the entire potted assembly is contained within
a metallic housing that acts as a magnetic shield to suppress
modulation of the photomultiplier tube gain as the external magnetic
field vector varies around the orbit (Rothschild et
al.\markcite{rer_spie} 1991).

\subsection{Particle Detectors}

Each HEXTE cluster includes three particle detection systems:  1) a set
of cosmic ray particle anticoincidence shield detectors, 2) an alpha
particle coincidence detector for each phoswich for gain control and
calibration, and 3) a single trapped radiation particle detector. The
anticoincidence shield system consists of four flat 0.64 cm thick
plastic scintillator modules configured in a four sided box around the
x-ray detectors.  The shields provide a prompt anticoincidence signal
for background events which are generated by energetic particle
interactions occurring in the external mass of the instrument and
satellite (Hink et al. 1991). The shields and the CsI(Na) of the
phoswich detectors are sensitive to minimum ionizing particles from
95\% of 4$\pi$ sr.

\begin{figure}[t]
\plotone{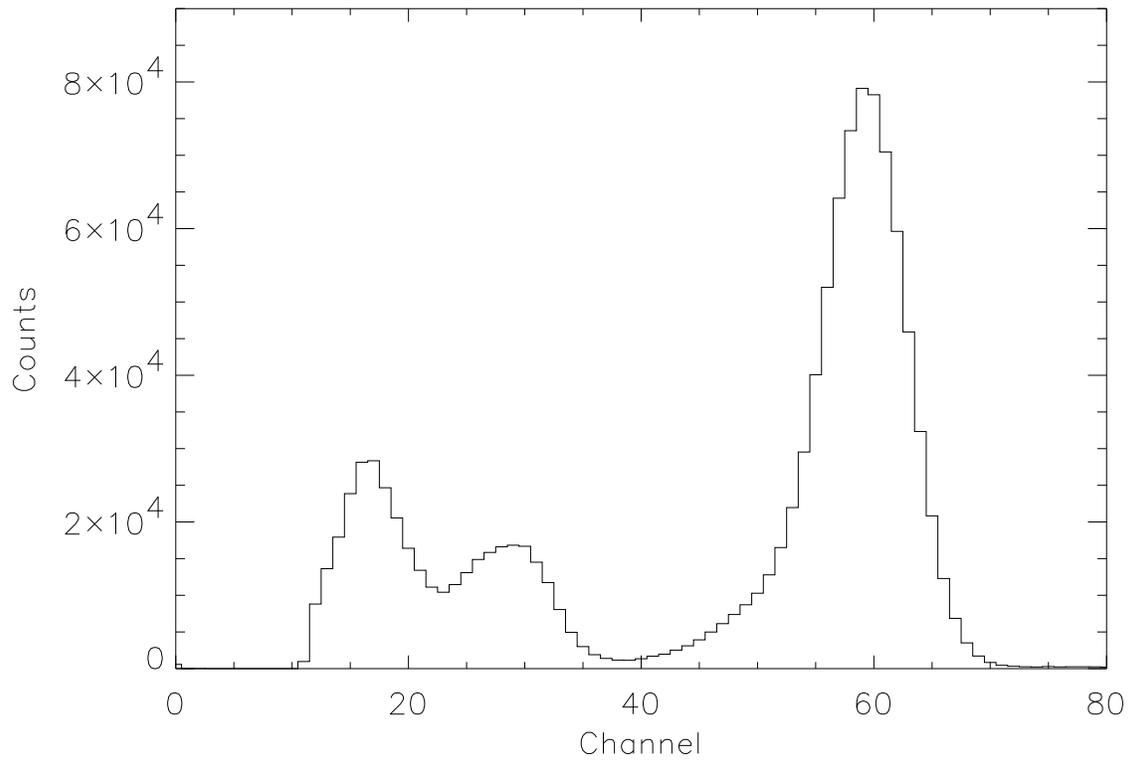}
\caption{Calibration histogram of the HEXTE $^{241}$Am gain
control source. The data were accumulated over one day on 10 January
1997. The 60 keV gamma-ray line, the complex of Np L de-excitation
lines around 17 keV, and the K-escape lines near 30 keV are evident.
Each channel corresponds to approximately 1 keV.\label{fig.calib}}
\end{figure}

Each phoswich detector module has an $^{241}$Am calibration x-ray
source mounted in the collimator immediately above its entrance window
and viewed by a separate 1/2 inch photomultiplier tube (see Figure
\ref{fig.detmodule}).  The calibration module provides continuous
feedback to the automatic gain control subsystem (Pelling et
al.\markcite{pelling} 1991), and calibration spectra (Figure
\ref{fig.calib}) are included in the flight telemetry. Gain control is
effected by varying the pulse height analyzer gain twice per second in
a feedback loop to keep the $^{241}$Am 59.5 keV line at a constant
pulse height. In-orbit performance shows that the line centroids are
stable to better than 0.02 PHA channels on a one day timescale, where
one channel is approximately 1 keV.  

The single particle monitor detector in each cluster is a 1.27 cm
diameter by 1.27 cm thick cylinder of plastic scintillator viewed by a 1/2 inch
photomultiplier tube. The aluminum housing provides a threshold energy
of $\sim$0.5 MeV for electrons. The particle monitors are used to
measure the ambient particle flux and safeguard the detector systems as
they pass through the trapped radiation belts of the South Atlantic
Anomaly (SAA). Safeguarding is accomplished automatically by reducing the
photomultiplier bias voltages by a factor of 4 during times of high
particle flux, thus limiting the anode and dynode currents to avoid
fatigue effects. The high voltage levels automatically return when the
particle monitor rate drops to a safe level.

\subsection{Source Beamswitching}

The HEXTE cluster rocking subsystem moves the viewing direction of the
detector arrays between on-source and off-source in order to obtain a
near-real time estimate of the instrument background.  The nominal
angular offset is 1.5$^{\circ}$ on either side of the on-source
position, with the option of increasing the offset to 3.0$^{\circ}$
and/or rocking to only one side in cases where there is a known
confusing source in a background field of view.  Since the rocking axes
of the two clusters are orthogonal to each other, four background
regions are nominally sampled around a given source position.  Data
acquisition is inhibited during the 2 s transitions between on- and
off-source positions, and the modulation of the two clusters is phased
such that one cluster is always viewing on-source.  The beamswitching
cycle consists of dwelling at the on-source position for 16, 32, 64, or
128 s, moving to an off-source position, dwelling there for the same
amount of time as on-source (less the 4 seconds of transit time),
moving back to the on-source position, and repeating for motion to the
other off-source position. This cycle is repeated throughout the
observation.

\subsection{HEXTE Electronics}

An 80C286 microprocessor running at 4.915 MHz is used to control each
cluster and process its output data.  Redundant interfaces are provided
from each cluster to the spacecraft Instrument Power Supply and
Distribution Unit (IPSDU) and 1773 fiberoptic command and telemetry
busses. The IPSDU also provides the conditioned and converted low
voltage power to the cluster and interfaces to the instrument rocking,
thermal control, and gain control subsystems. The electronics on each
cluster exchange SAA detection and burst detection signals with the
other, and, in the case of burst detection, exchange such
signals with the PCA's data processor, known as the Experiment Data
System (EDS).

\subsection{Data Processing/Telemetry Formatting}

In-flight programmable logic is used to control data selection on an
event by event basis. Data for each event that passes the selection
criteria are combined with other subsystem indicators and collected to
form a 56-bit event code. This
event code is then passed on to either the automatic gain
control/calibration subsystem or the science processing/telemetry
formatting process.  The default data selection criteria are for the
event energy loss to be within the HEXTE energy range, the event pulse
shape to be that of pure NaI with no contribution from CsI, no recent
(within 2.5 ms) large energy loss ($>$ 20 MeV), and no anticoincidence
shield pulse present. This provides
efficient data recovery for most observations by measuring the energy
loss for only NaI(Tl) events. The HEXTE energy range is determined by the
settings of the electronic pulse height discriminators. The lower level discriminator is
commandable between 5 and 50 keV, and in practice the default value is
12 keV. The upper level discriminator is fixed at 250 keV. System noise
causes these discriminator edges to be imperfect, and the non-zero energy
resolution results in some loss of events at the energy extremes.
Consequently, data analysis is recommended to be in the 15$-$250 keV
range.

Two standard mode data products are generated within HEXTE every 16
seconds.  The purpose of these ``Archive''  modes is to provide a basic
temporal and spectral record of every source observed that is
independent of the scientific mode chosen by the Guest Observer.  The
Archive Histogram mode produces a pseudo-logarithmically compressed 64
channel pulse height histogram for each phoswich every 16 seconds. The
Archive Multiscalar mode produces four light curves from the sum of all
four phoswiches in a cluster with 1 second temporal resolution. The
four light curves represent the approximate energy ranges of 15 to 29,
30 to 61, 62 to 125, and 126 to 250 keV. A measurement of the livetime
for each phoswich for each 16 s Instrument Data Frame (IDF) is included
with the standard mode data.

HEXTE science processing in each cluster can be configured to produce one of three
modes at any given time: 1) Event List mode, 2) Histogram Bin mode, or
3) Multiscalar mode. The Event List mode is an event-by-event list of a
commandable subset of the event code for each photon detected. Any
combination of the 7 bytes comprising the event code may be selected
for inclusion in telemetry.  Subsequent ground analysis is able to
generate light curves and/or spectra, depending upon the event code
bytes selected.  The Histogram Bin mode produces a pulse height
histogram for each detector, or, alternatively, for the sum of all four
detectors in a cluster. The histograms can be tailored to each
observation by choice of the number of histograms per IDF, the number
of pulse height bins, the count capacity of each bin, and the full
scale pulse height channel value of each histogram.  The Multiscalar
Bin mode produces up to eight light curves, contiguous in energy with
channel boundaries fully selectable, with commandable time sampling
from 0.5 ms to 1 second, for either each detector or the sum of all
four. A ninth light curve is available to accumulate the Lost Events
(i.e., otherwise valid events that were not pulse height analyzed due
to the analyzer being busy with a previous event) on the same time
scale as the other light curves.

A fourth mode, Burst List, can be enabled by command to run in parallel
with any of the three basic science modes. This mode takes a
``snapshot'' of the data from a bright source at the highest spectral
and temporal resolution for later transmission.  The Burst List mode
buffers 4 bytes of the event code from each event in a circular buffer
containing 25,600 events.  Upon initiation by a burst trigger, the
buffer accumulates a software commandable number of post-trigger
events, then freezes its contents, and awaits a command to download its
contents in place of the normal science data. Deadtime data are
accumulated once each second for the first 48 seconds after the burst
trigger, and included in the Burst List buffer.  A burst trigger can be
generated by any of four sources: a serial digital ground command, a
trigger received from the EDS electronics, a trigger received from the
other cluster, or an internally calculated burst trigger.  The internal
burst trigger is issued upon detection of an increase in the recent
average event rate exceeding a commandable threshold.

\section{Instrument Performance}

Two anomalies have occurred which affect HEXTE performance:
1) the pulse height analyzer of one detector has failed, and 2) the
system for measuring deadtime within the instrument does not properly
register deadtime after large energy losses in the crystals. On 6 March
1996 the pulse height analyzer for the third phoswich in cluster B
stopped digitizing  pulse heights. The most probable cause was
the failure of a single element, such as an operational amplifier or
solder connection. Rates from this detector remain nominal, but 
energy information and the gain control function are lost.

During analysis of IOC data, it was realized that the analog circuitry
that processes pulse shape information for individual x-ray events was
paralyzed after large energy losses for times substantially longer than
anticipated. Since this system is required to identify NaI energy
losses for subsequent pulse height analysis, the consequence of this
problem is to have the system reject otherwise valid events during the
"recovery" period without incrementing the internal deadtime counters
as intended.  Consequently, the hardware deadtime associated with a
large energy loss was increased from 0.3 to 2.5 ms to minimize this
effect. In addition, the remaining unmeasured deadtime has been modeled
using parameters associated with the $>$20 MeV extreme upper level
discriminator rate (XULD) and to the $>$250 keV upper level
discriminator rate (ULD).  These two parameters were determined for
each detector by requiring the flux from the Crab Nebula to be constant
on a 16 second timescale.  Based upon observations of the Crab
Nebula/Pulsar spectrum (see below) the deadtime appears to be mostly
energy independent and only the normalization of measured fluxes is
affected.  The total mean HEXTE corrected deadtime averages about
40\%.

\begin{figure}
\plotone{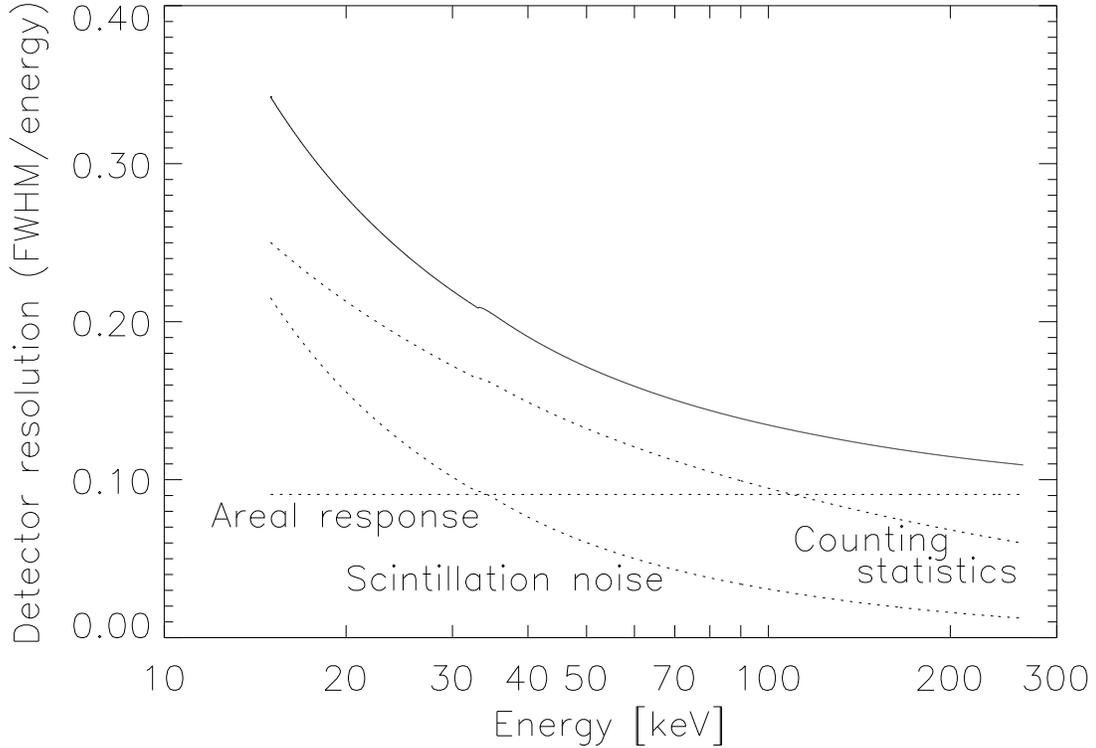}
\caption{The energy resolution, $\Delta$E/E, as a function of
energy, E, for one of the HEXTE detectors. The three components of the
energy resolution are shown. The component labeled ``Counting
statistics'' is due to the stochastic nature of the charge generation
in the photomultiplier tube and has an E$^{-0.5}$ dependence. The
constant factor labeled ``Areal response'' reflects the variation of
light collection efficiency across the detector. Finally, the
additional width due to long lived states in the NaI(Tl) stimulated by
cosmic rays contributes proportional to E$^{-1}$, and is labeled
``Scintillation noise''. The small bump at 33 keV results from the
lower light output of NaI(Tl) for events just above the K-edge of
iodine due to the lower energy of the
photoelectron.\label{fig.resolution}}
\end{figure}

\begin{figure}
\plotfiddle{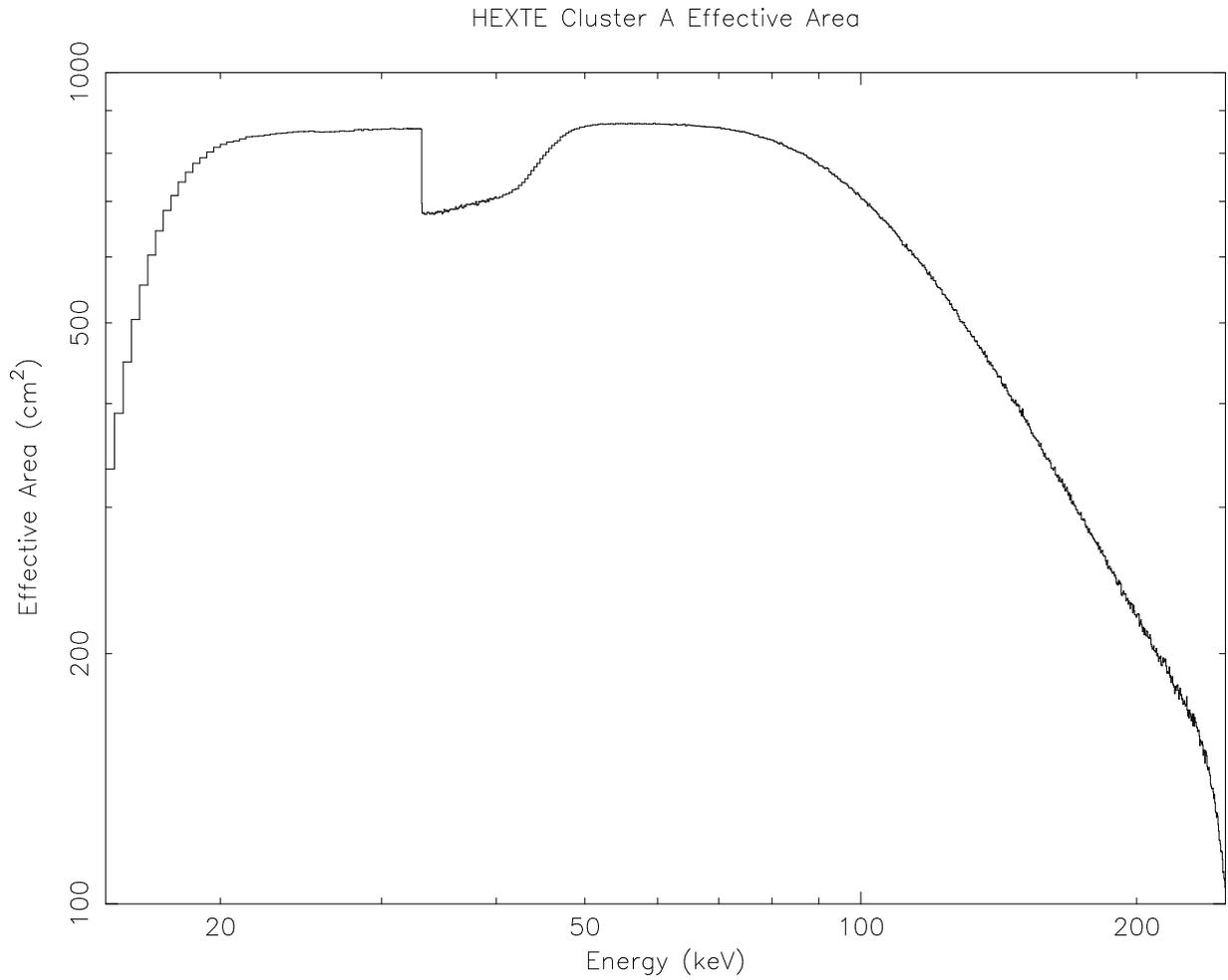}{6in}{270.}{70.}{70.}{-252}{472}
\caption{The HEXTE effective area versus energy for cluster A. The
reduction at lower energies is due to photoelectric absorption in the
housing above the detector, while the reduction at higher energies is
due to the finite NaI(Tl) thickness. The sharp drop and smooth recovery near 30 keV is
due to the change in photoelectric cross section at the K-edge of
iodine and the onset of K-escape losses.\label{fig.area}}
\end{figure}

The average energy resolution of the eight detectors was seen to
increase slightly in orbit from the prelaunch average of 14.8\% to
15.4\% at 60 keV (Hink, Pelling, and Rothschild\markcite{hink_etu}
1992). This increase is believed to be due to presence of additional
``noise'' in the scintillation light signal due to long-lived decay
states that are excited by large cosmic ray energy loss events in the
NaI(Tl) crystals. The energy resolution of the phoswiches is given in
Table 2, and shown as a function of energy in Fig.
\ref{fig.resolution}.  The HEXTE effective area as a function of energy
is shown in Fig.  \ref{fig.area}.  This incorporates the estimated
physical open area per detector and the energy-dependent response of
the detector as determined by Monte Carlo simulation. The simulation
reproduces data taken before launch using both radioactive sources and
monochromatic x-rays at the Brookhaven National Synchrotron Light
Source (Wayne et al. \markcite{len} 1997). It also includes the effects
of K-escape x-rays, Compton scattering, the non-linearity of light
output of NaI(Tl), and  pulse shape discrimination versus energy.

\subsection{Instrument Background}

\begin{figure}[t]
\plotone{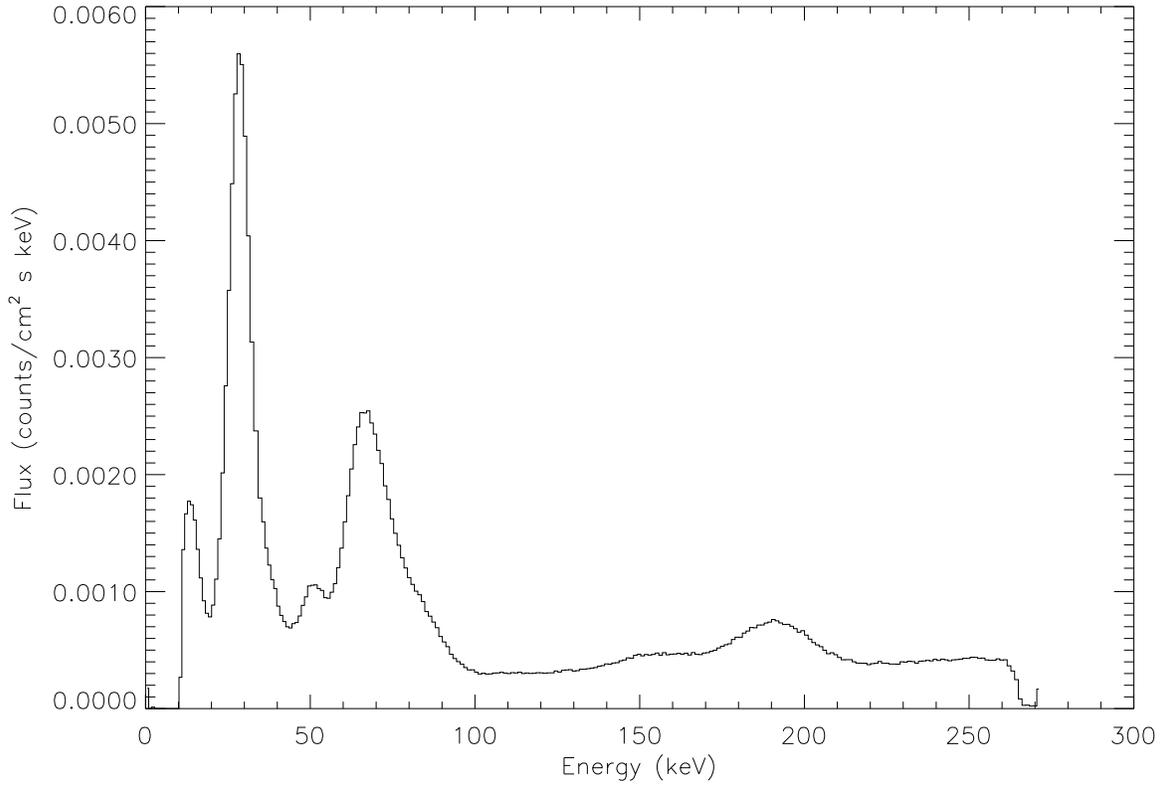}
\caption{The HEXTE background for a single detector plotted versus
energy.  Background lines due to activation of the iodine by cosmic and
trapped radiation are evident at 30, 55, 66, and $\sim$190 keV, as are
the K$_{\alpha \beta}$ lines at 74 and 85 keV from fluorescence of the
lead collimator. The data are from a 7.5 hour observation of blank sky
fields.\label{fig.backgrnd}}
\end{figure}

Figure \ref{fig.backgrnd} shows the HEXTE background from 7.5 hours of
observing blank fields on January 15, 1996 during the In-Orbit Checkout
phase of the mission.  The background is dominated by internal
activation effects, with less than 5\% of the total due to the cosmic
x-ray background transmitted by the 1$^\circ$ aperture of the
collimators. Prominent background lines due to activation are seen at
30 keV (the K lines from the Te daughters of various iodine decays), 55
keV (due to $^{121}$I decay), and near 190 keV (due to $^{123}$I). K
escape lines from the 190 keV complex are also evident at $\sim$160
keV. The blend of K lines from fluorescence of lead in the collimator
by charged particles and cosmic x-rays is prominent from 60 to 90 keV.
The instantaneous, 15-250 keV background rate varies between 80 and 110
count/s in each cluster depending upon location in the orbit.  This
translates to 4.3 to 5.9 $\times$ 10$^{-4}$ count cm$^{-2}$ s$^{-1}$
keV$^{-1}$ averaged over the full energy range.

\subsection{Orbital Environment}

\begin{figure}[t]
\plotfiddle{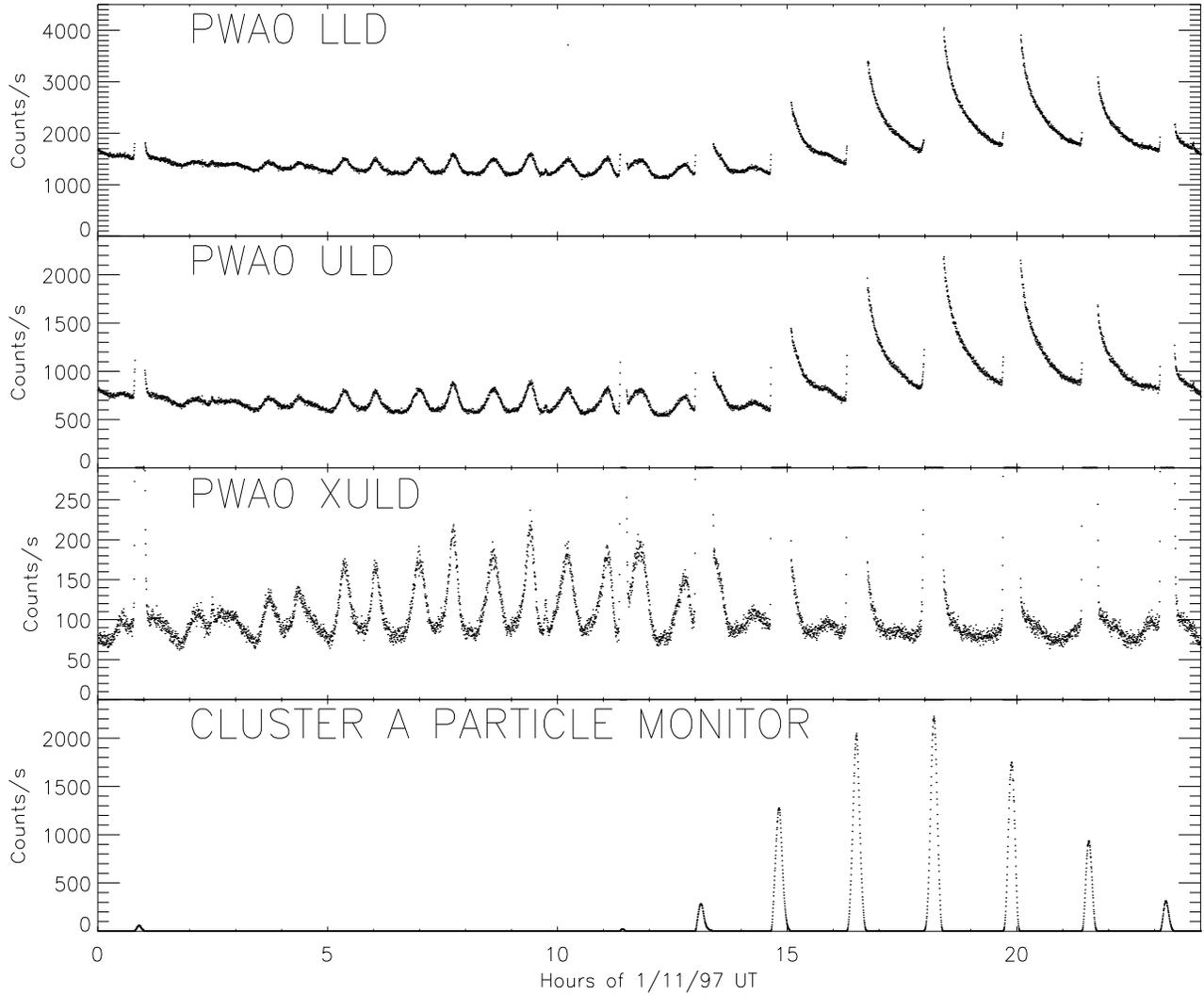}{6in}{90.}{75.}{80.}{288}{0}
\caption{Counting rates from one HEXTE detector (Phoswich 0 in
Cluster A) for 11 January 1997. The rates from the phoswich
detector are:  the Lower Level Discriminator (LLD, events with energy
losses $>$ 12 keV), Upper Level Discriminator (ULD, events with energy
loss $>$ 250 keV), and eXtreme Upper Level Discriminator (XULD, events
with energy loss $>$ 20 MeV). The fourth panel shows the Particle
Monitor rate (charged particles with $>$ 0.5 MeV energy loss) which
experiences large increases when transiting the SAA.\label{fig.rates}}
\end{figure}

The {\it RXTE} orbit of 580 km altitude and 23$^{\circ}$ inclination
results in the satellite passing through the trapped radiation of the
South Atlantic Anomaly (SAA) for about one-half of the orbits in a day
and sampling a range of geomagnetic latitude throughout the day. Figure
\ref{fig.rates} shows typical counting rates  from one HEXTE detector
over a day. The Lower Level Discriminator (LLD) rate is the total
counting rate in the scintillators above 12 keV.  Modulation of the
charged particle flux due to varying geomagnetic rigidity cutoff is
apparent in the non-SAA orbits between hours 3 and 13. Shutdown of the
detectors during SAA passages and the subsequent 25 minute half-life radioactive
decay of $^{128}$I upon exit of the SAA are evident between hours 13 to
24.  The Upper Level Discriminator (ULD) rates reflect the total
counting rate in the scintillators above 250 keV, and is similar to the
LLD rates apart from normalization. The eXtreme Upper Level
Discriminator (XULD) rate monitors events leaving more than 20 MeV in
the crystals. This rate is dominated by interactions of
energetic cosmic rays and exhibits modulation in the non-SAA orbits.
Typical counting rates for the particle monitor (which are dominated by
trapped radiation of the SAA) are also shown in Figure \ref{fig.rates}.

\subsection{Background Subtraction}

Since the HEXTE counting rate exhibits temporal variations in any given
orbit, as seen in Figure \ref{fig.rates},  one must make a near real
time estimate of the background utilizing the source chopping system
described above.  This method's ability to provide accurate background
subtraction is essential to HEXTE sensitivity. 

\begin{figure}
\plotfiddle{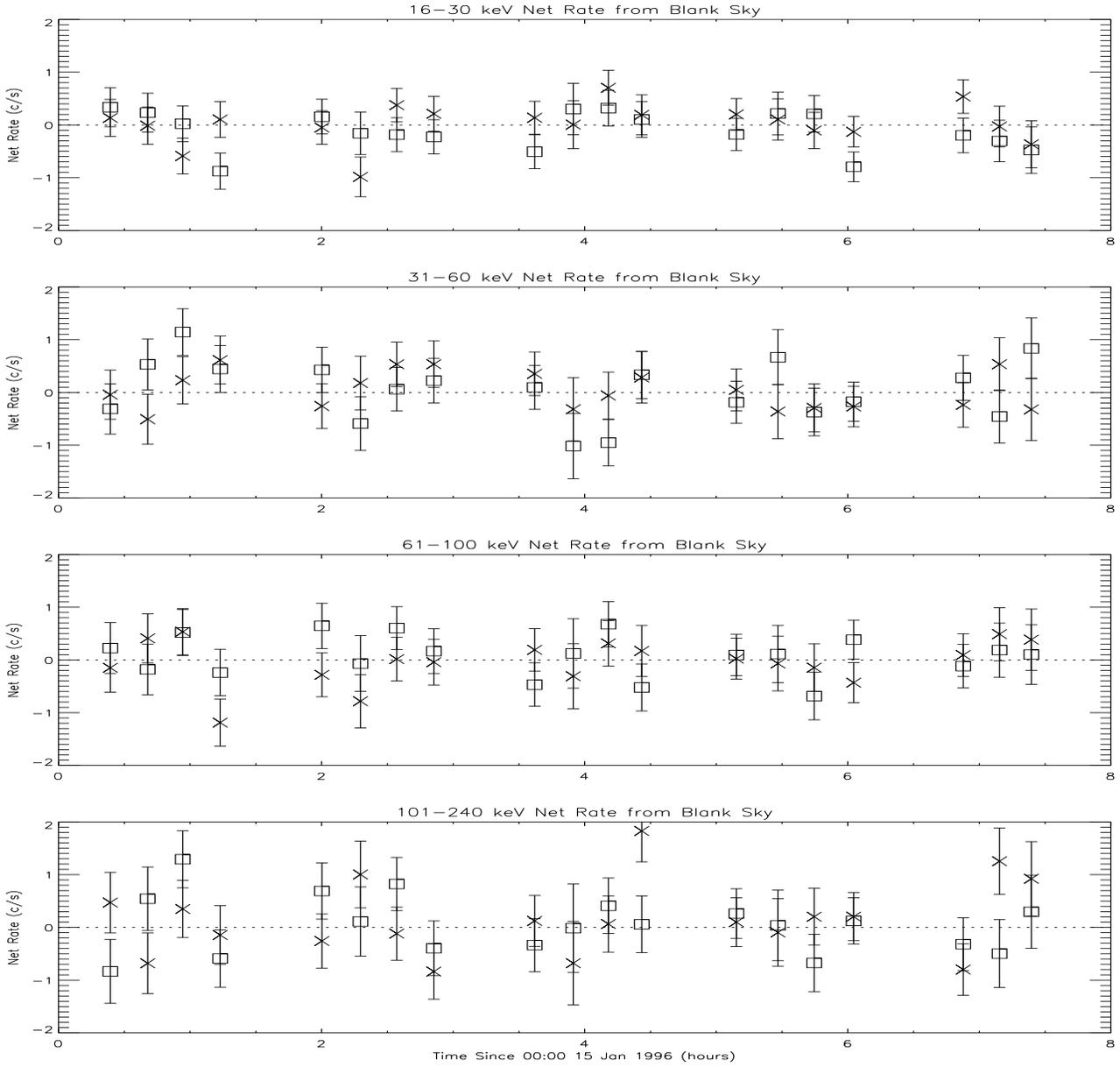}{6in}{0.}{100.}{65.}{-324}{0}
\caption{Background subtracted counting rates for 19 ten-minute
observations of blank fields on 15 January 1996. The four sets of rates
cover the energy range 16-30, 31-60, 61-100, and 101-240 keV. The crosses
(x) represent Cluster A and boxes ($\Box$) Cluster B. The dotted line
indicates zero net flux.\label{fig.4rates}}
\end{figure}

One test of the background subtraction is to compare net rates to zero
when viewing blank fields. The blank sky observations of 15 January
1996 covered 19 different pointing directions determined to be source
free (flux $<$2 PCA c/s assuming a power law index of 1.7) from the
ROSAT WGACAT (White, Giommi, Angellini CATalog; White, Giommi, and
Angellini \markcite{wgacat} 1994). These observations averaged $\sim$10
minutes each.  Fig. \ref{fig.4rates} shows the background subtracted
rates in four energy bands (16$-$30, 31$-$60, 61$-$100, and 101$-$240 keV) for
each cluster for each pointing direction.  An average net rate and
uncertainty was calculated for each rate and the chi-square calculated
for comparison to the average net rate.  Table 3 gives these values
along with the probability of exceeding chi-square by a random
fluctuation.  In all cases the probability was less than 8\%,
indicating that the background subtraction method used by HEXTE is
quite effective for short observations.

\begin{figure}[t]
\plotfiddle{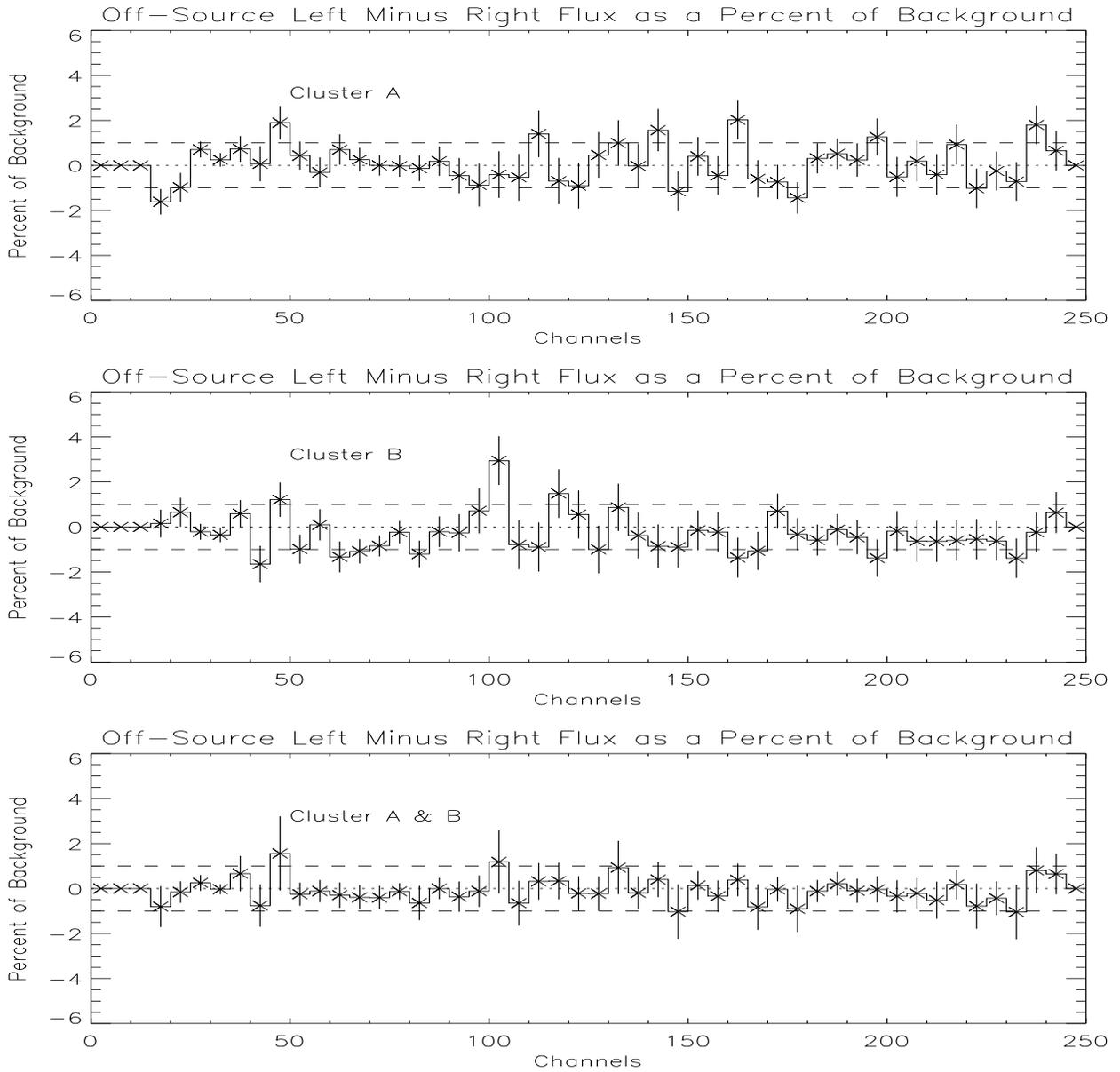}{6in}{0.}{100.}{65.}{-324}{0}
\caption{The net counting rate histogram of the difference of
the two off-source positions for each cluster and their sum
during the MCG 8-11-11 observation. One percent of the total off-source
flux is shown by pairs of horizontal dashed lines for comparison.
``Left'' and ``Right'' correspond to the plus and minus directions
of the off-source viewing.\label{fig.mcg_lr}}
\end{figure}

Another test is to compare the spectra of the two off-source positions
in a long observation that experiences a variety of orbital
conditions.  In January of 1996 a 200 ks observation of MCG 8-11-11 was
performed (see below) with 1.5$^{\circ}$ offsets and 32 s dwell times.
Background data from each off-source position were accumulated for both
clusters. Subtraction of one blank sky off-source region from another
provides a measure of the accuracy of this technique for long
observations, where it is most important. The 200 ks observation
resulted in 46 ks of on-source livetime for each cluster (92.5 ks of
800 cm$^2$ total) and 40 ks at each off-source position for each
cluster (80.6 ks of 800 cm$^2$ total).  The net flux from these two
off-source, blank sky positions is shown in Figure \ref{fig.mcg_lr}.
Binned every 5 pulse height channels, fluctuations from bin to bin are
$\ltsim$1\% of background, and the broadband average is
(0.11$\pm$0.10)\%, (-0.34$\pm$0.10)\%, and (-0.11$\pm$0.07)\% for
cluster A, cluster B, and their sum, respectively.  At 100 keV the
background is 3$\times$10$^{-4}$ count cm$^{-2}$s$^{-1}$keV$^{-1}$. For
equal on- and off-source viewing times of 10$^5$s, this implies a
3$\sigma$ minimum detection sensitivity in a 20 keV band at 100 keV of
3$\times$10$^{-6}$ photons cm$^{-2}$s$^{-1}$keV$^{-1}$, or 1\% of
background (Gruber et al.\markcite{gruber} 1996). This shows the HEXTE
background measurement technique is extremely effective for typical,
long, weak source observations.

\begin{figure}
\plotfiddle{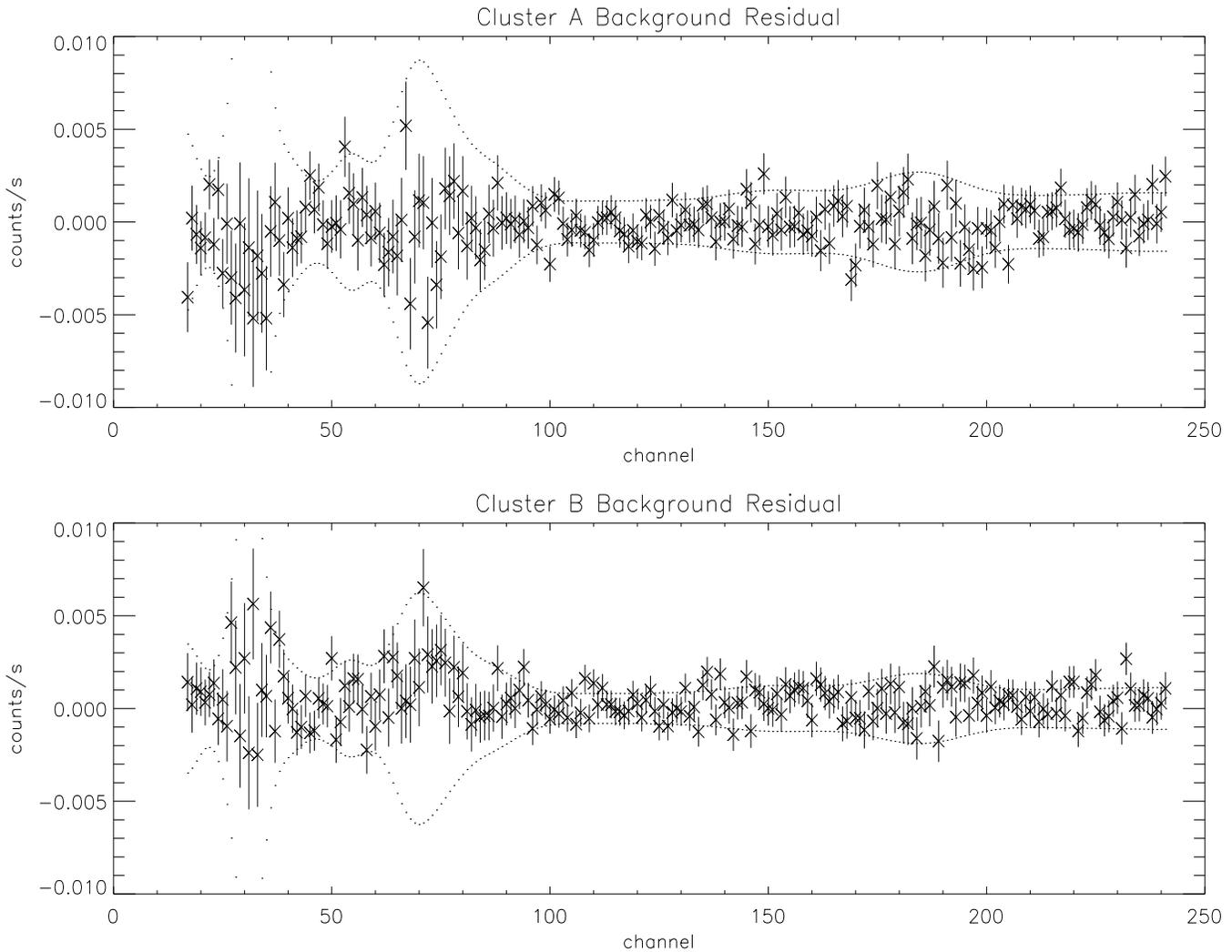}{6in}{90.}{75.}{80.}{288}{0}
\caption{The net off-source counting rate histogram versus pulse
height channel for a 5$\times$10$^6$ s accumulation of observations
with greater than 50 ks duration for the two HEXTE clusters. The
detector gain is approximately one keV per channel, and the energy
range covered is 15$-$240 keV. The dotted lines represent $\pm$0.5\% of
the background per channel.\label{fig.dannyboy}}
\end{figure}

Finally, analysis of the background level as a function of aperture
viewing direction for all observations of 50ks or more is part of a
doctoral thesis at UCSD on the diffuse background. This has led to a
spectrum of the difference between the two off-source viewing
directions (Fig. \ref{fig.dannyboy}) accumulated over more than
5$\times$10$^6$ s. (The on-source data are not available to us due to
the proprietary nature of the observations.) The largest, single
channel deviation is $\leq$1\% of the background rate in that channel
(MacDonald et al.\markcite{danmac} 1996). The 15$-$240 keV net rates in
clusters A and B are (-0.2$\pm$1.5)$\times$10$^{-3}$ and
(0.5$\pm$1.2)$\times$10$^{-3}$ counts s$^{-1}$.  The chi-square fit of
the 15$-$240 keV data to these averages was 211.0 and 215.1 for 224
degrees of freedom, respectively.  This establishes that long term
integrations with HEXTE have systematic spectral uncertainties at a few
tenths of a percent of background.

\subsection{Spectral Calibration}

\begin{figure}
\plotfiddle{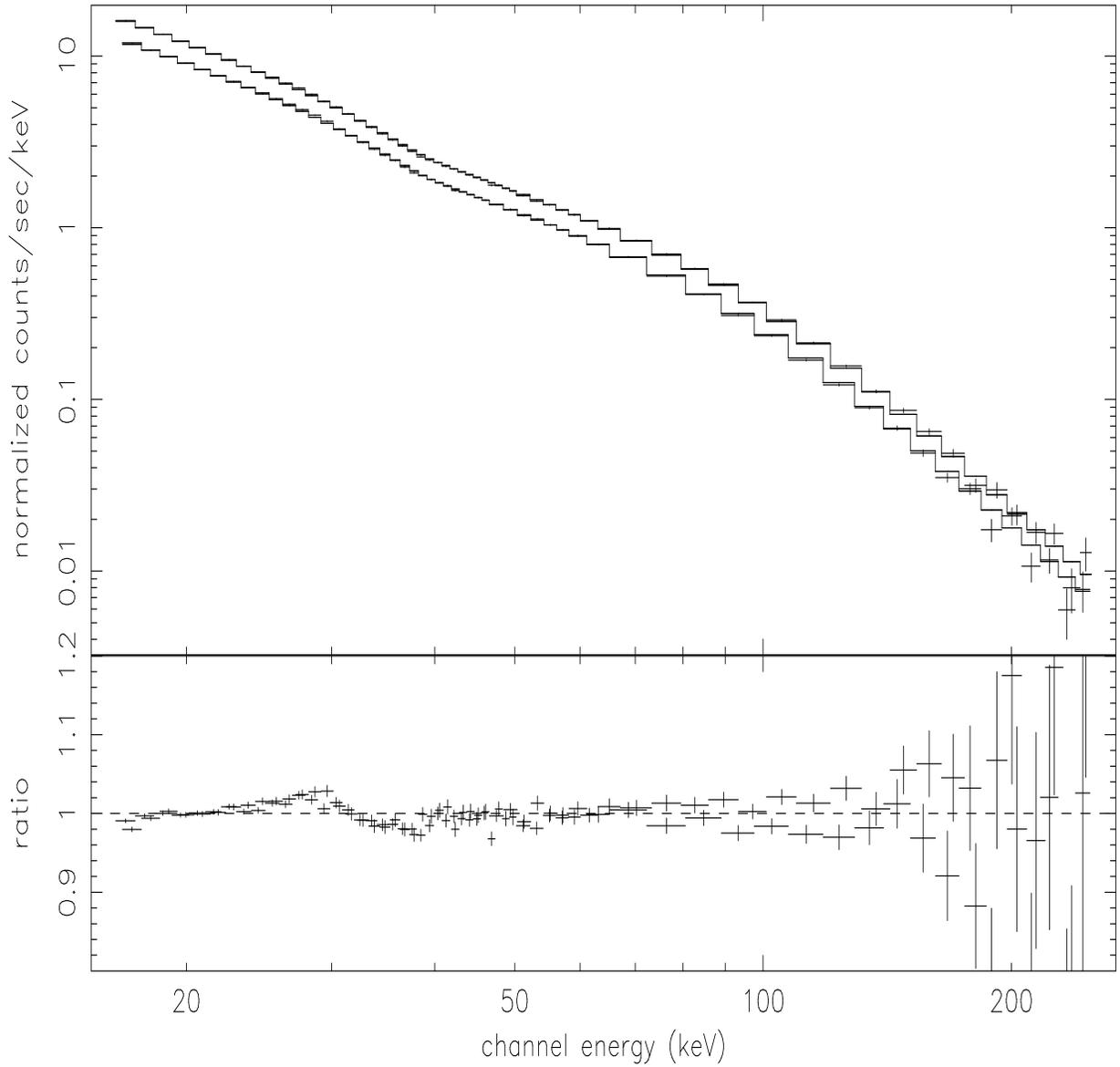}{6in}{270.}{70.}{100.}{-252}{544}
\caption{The HEXTE counts histogram versus energy plus best fit
broken power law model for fitting the Crab Nebula/Pulsar spectrum in
Clusters A and B simultaneously. The upper plot shows the observed data
plus model for each cluster with higher energy data rebinned into 10
keV groups for display only. The lower plot shows the ratio of the
residuals of the best fit model to the data.  \label{fig.crab_spec}}
\end{figure}

Using the HEXTE effective area, which includes our present best estimate
of the calibration parameters, we have fitted the total, dead time
corrected, Crab Nebula/Pulsar spectral data from each HEXTE cluster in
the 15$-$240 keV range. The best fit photon index for a single power
law was $\Gamma$=2.062$\pm$0.001 with a normalization at 1 keV of
6.73$\pm$0.02 photons cm$^{-2}$s$^{-1}$keV$^{-1}$.  The ratio of this
normalization between clusters A and B was 1.023. This observed index
is somewhat flatter than seen by other instruments (Bartlett
\markcite{bartlett} 1994 and references therein).  Since the spectrum
is known to steepen near 100 keV (Jung \markcite{jung} 1989), the value
of an effective average index is dependent on the energy range over
which the fit is attempted.  Accordingly, the HEXTE data were fit to a
power law that breaks sharply at a single energy to a second power law
(i.e., a broken power law model). In this case the photon index below
the break energy was found to be $\Gamma_1$=2.045$\pm$0.002, the break
energy was E$_B$=57.2$\pm$3.0 keV, and the photon index above the break
energy was $\Gamma_2$=2.140$\pm$0.009. The chi-square was reduced by
210 for 433 degrees of freedom in going from the single power law to
the broken power law. The reduced chi-square for the latter was 1.80.
The instrument counts histogram and best fit broken power law model are
shown in Fig. \ref{fig.crab_spec} along with the ratio of the data to
the model. The higher energy data have been rebinned into 10 keV groups
for display only --- the fitting was done to 219 pulse height channels
in each cluster.

The broken power law fits can be compared to Barlett
\markcite{bartlett} (1994), who reported $\Gamma_1$=2.00$\pm$0.03,
E$_B$=60$\pm$7 keV, and $\Gamma_2$=2.22$\pm$0.03, and to Jung
\markcite{jung} (1989) with $\Gamma_1$=2.075$\pm$0.005,
E$_B$=127.7$\pm$3.6 keV, and $\Gamma_2$=2.48$\pm$0.03. The HEXTE broken
power law normalization at 1 keV was 6.38$\pm$0.03 photons
cm$^{-2}$s$^{-1}$keV$^{-1}$, which is less than seen for HEAO-1
(8.76$\pm$0.03 photons cm$^{-2}$s$^{-1}$keV$^{-1}$; Jung
\markcite{jung} 1989) in the 17$-$2300 keV range.  The majority of this
normalization discrepancy, we believe, is in the assumed shape of the
collimator response near its peak. A perfect, pointed peak is assumed
presently. The more accurate collimator shape is being calculated from
multiple scans across the Crab, and will be the basis for a revision in
the HEXTE open area calculation in the near future. We have found
minimal energy dependence to the normalization
($\Gamma_1$=2.050$\pm$0.002 and 2.037$\pm$0.003 for clusters A and B
respectively), and thus, fitting with a normalization parameter as part
of the fitting procedure for other sources is warranted.  These details
are the subject of continuing calibration efforts with the PCA and
HEXTE instrument teams. The differences between the HEXTE fitted values
and those from HEAO-1 and the GSFC balloon instrument reflect the
present HEXTE systematic uncertainties, those of those past missions,
and the fact that a broken power law is too simple of a model for the
spectral evolution in the Crab total spectrum.

\subsection{Absolute Timing}

\begin{figure}[t]
\plotfiddle{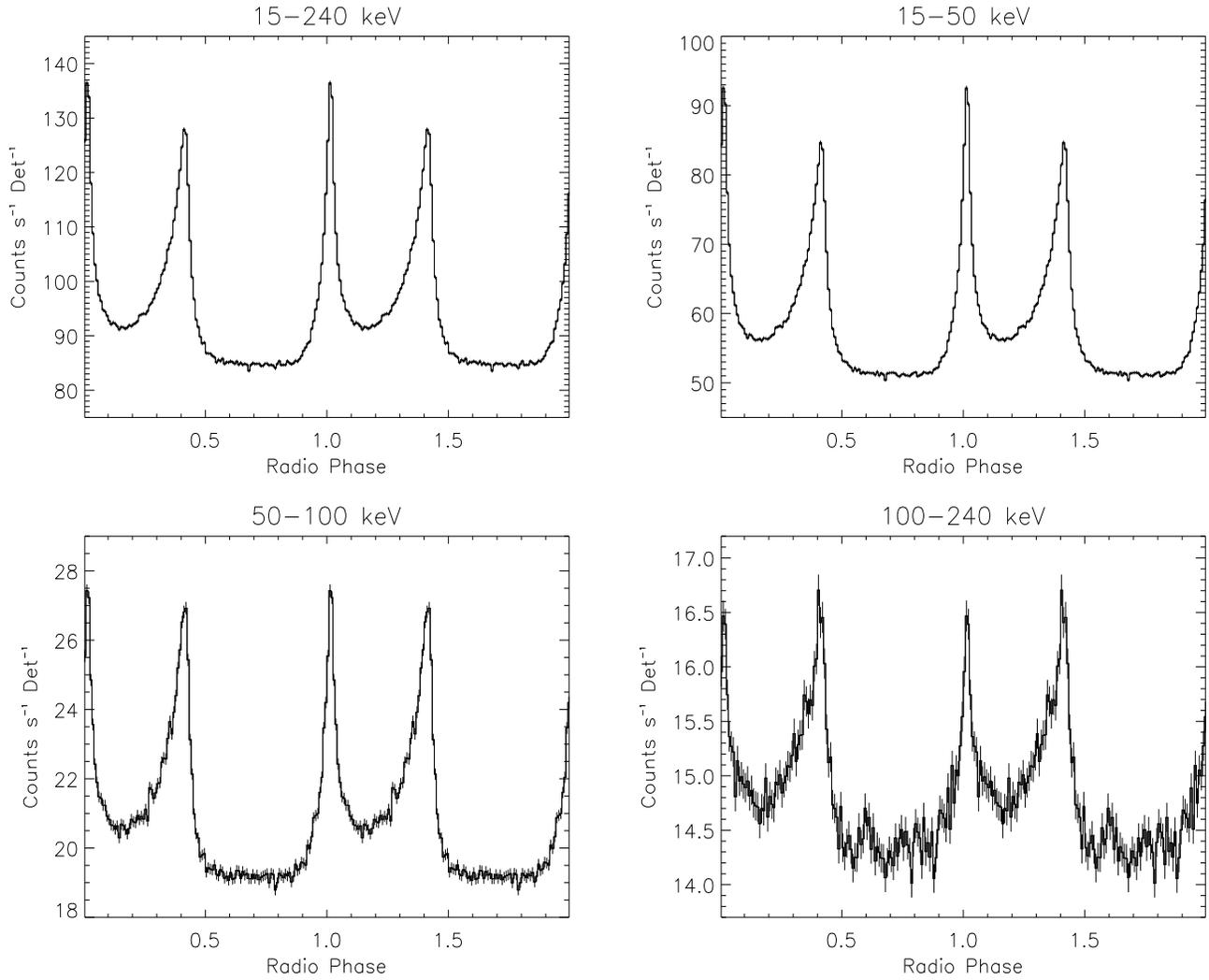}{6in}{90.}{75.}{80.}{288}{0}
\caption{Folded light curves of the Crab Nebula and Pulsar based
upon the radio ephemeris. Light Curves are shown for the entire 15$-$250
keV HEXTE energy range and for the energy ranges 15$-$50 keV, 50$-$100 keV,
and 100$-$240 keV. The x-ray to radio offset is less than 1 ms.\label{fig.crab}}
\end{figure}

The HEXTE timing information, as well as that of {\it RXTE}, was
verified through observations of the Crab pulsar. The arrival time of
each photon was corrected to the solar system barycenter and folded using the
Crab pulsar radio ephemeris appropriate to that time. Figure \ref{fig.crab}
shows the light curves for the entire HEXTE 15$-$250 keV energy range and
for three energy intervals within the HEXTE range. The pulse profile
consists of two sharp peaks separated by $0.4$ in phase but connected
by a ``bridge'' of pulsed emission. The sharp pulsed components plus
the bridge, or interpulse, connecting them are as expected from
previous observations (e.g. Ulmer et al.\markcite{ulmer} 1994). Zero
phase is defined as the center of the first peak in the radio with the
ephemeris parameters given in Table 4. For this observation the first
peak is within one phase bin (300 $\mu$s) of the radio peak, and thus
the HEXTE absolute time reference is accurate within a fraction of a
millisecond. From the plot it is apparent that the second peak in the
folded light curve has a harder spectrum than the first peak, as
indicated by its increasing dominance over the first peak with
increasing energy.

\subsection{Response to High Rates}

\begin{figure}[t]
\plotfiddle{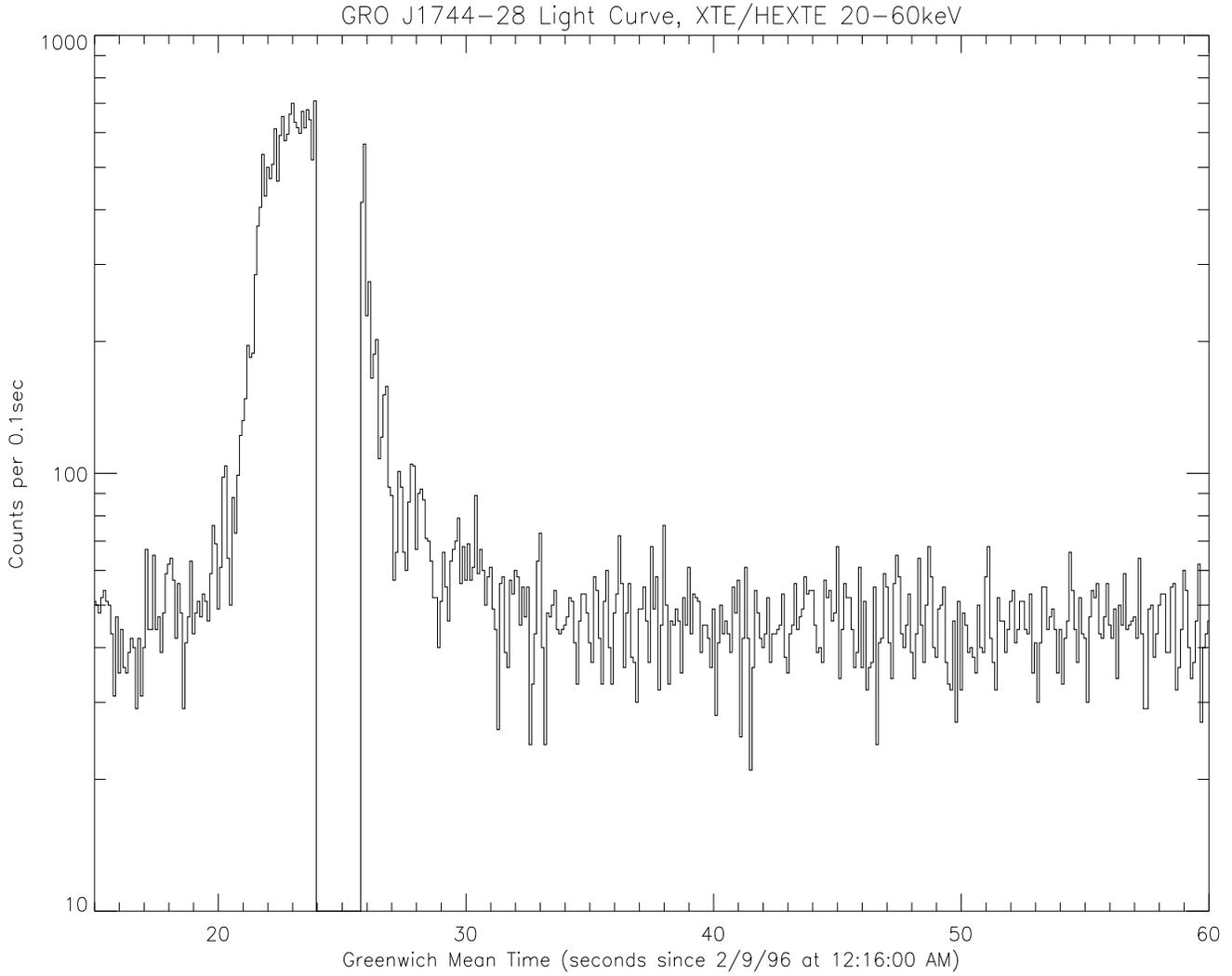}{6in}{90.}{75.}{80.}{288}{0}
\caption{The light curve of the bursting pulsar 1744-28 taken
during the In Orbit Checkout phase of the {\it RXTE} mission.
Individual pulses at the 2.1 Hz pulse period are evident in the
persistent flux and a burst is also evident. The $<$2 s dropout and
recovery of data during the burst were caused when x-ray rates exceeded
the maximum transferable rate.\label{fig.1744}}
\end{figure}

Three weeks after launch, {\it RXTE} began a series of Target of
Opportunity observations of the bursting pulsar, GRO J1744-28.  At this
time the source's 2$-$10 keV flux was about twice that of the Crab with
bursts up to a factor of ten brighter lasting for about 10 s. The HEXTE
was designed to handle persistent rates as high as 5 times that of the
Crab, and to recover quickly from rates in excess of that. The bursts
from GRO 1744-28 clearly exceeded that rate.  This is demonstrated in
Figure \ref{fig.1744} where the 2.1 Hz pulsations are seen before and
after the $<$2 s of lost data at the peak of the burst. HEXTE lost data
at the peak of the counting rate due to a finite number of data
buffers, and the data transmission resumed once buffers again became
available.

\subsection{Hard X-ray Spectrum of a Weak Source (MCG8-11-11)}

The Seyfert 1 active galaxy MCG8-11-11 was observed over the period
from 7--29 January 1996 to verify the overall performance of HEXTE when
observing a weak source. Basic results on the 15$-$240 keV HEXTE spectrum
of MCG 8-11-11 are presented to demonstrate the HEXTE capabilities for
spectral studies of weak sources, such as active galaxies. Further
analysis and interpretation of MCG 8-11-11 results will be the topic of
a future paper.

This source has been observed in
the 0.5$-$20 keV range over the last 25 years (Elvis et
al.\markcite{elvis} 1978; Mushotzky et al.\markcite{mushy} 1980; Petre
et al.\markcite{petre} 1984; Turner \& Pounds\markcite{turner} 1989;
Treves\markcite{treves} et al. 1990; Grandi et al.\markcite{grandi}
1997) to be a simple power law ($\Gamma \approx$1.7) modified by low
energy absorption (N$_H \approx$2.5$\times$10$^{21}$ cm$^{-2}$). At
higher energies, the OSSE measurement (Kurfess, Johnson, \&
McNaron-Brown\markcite{kurfess} 1997) found a steeper index
($\Gamma$=2.7$\pm$0.4) in the energy range 50$-$150 keV, and the MISO
balloon instrument (Perotti et al.\markcite{perotti} 1981) found a
flatter ($\Gamma$=1.0$\pm$0.7) index from 20 keV to 1 MeV.

\begin{figure}[t]
\plotfiddle{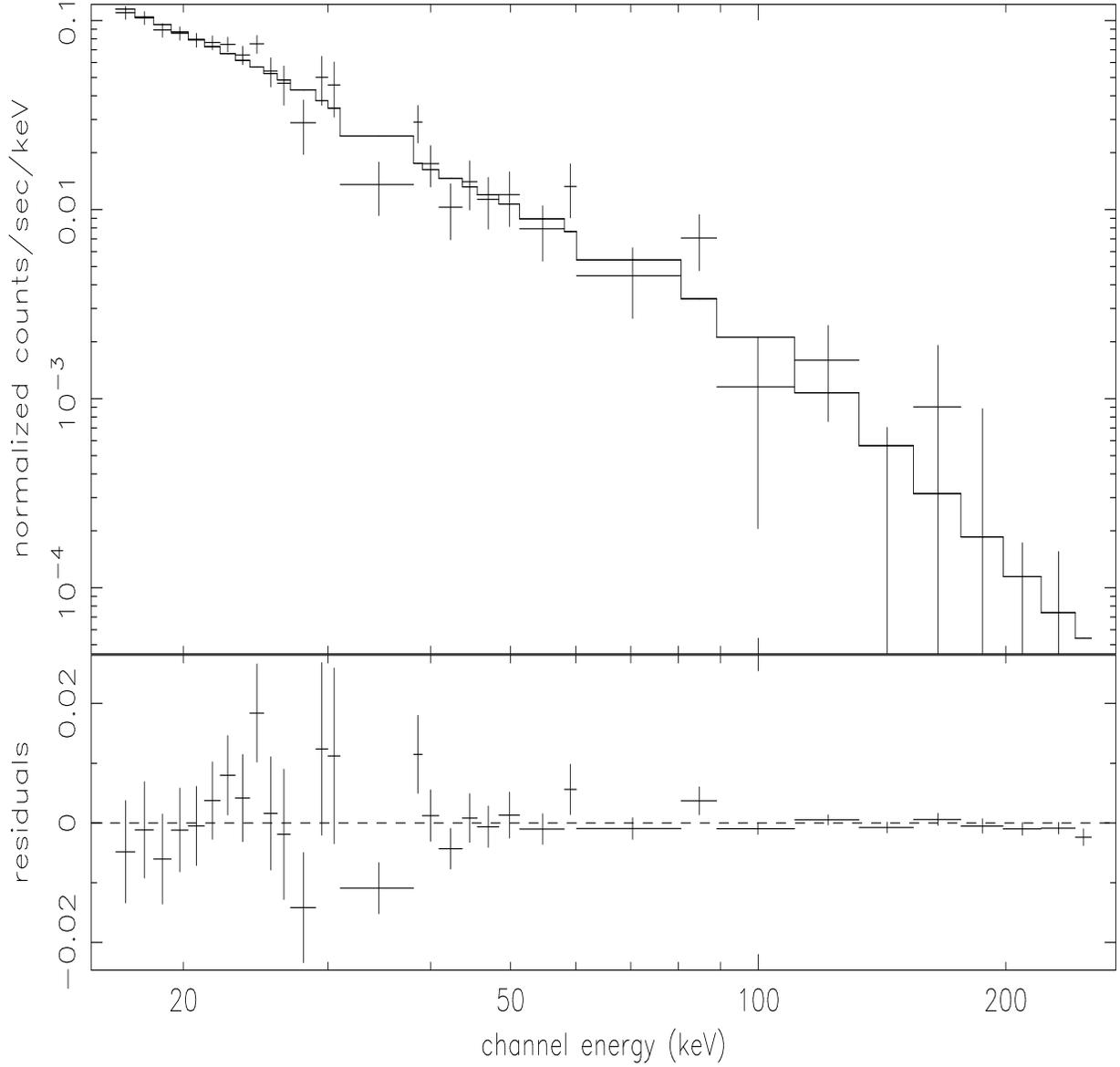}{6in}{270.}{70.}{100.}{-252}{544}
\caption{The observed HEXTE counts histogram from MCG8-11-11
compared to the best-fit power law model.  The HEXTE data at higher energies have
been rebinned for display only into $\sim$20 keV bins.  The lower panel
displays the residuals of the observed counts to the predicted best-fit
model counts.\label{fig.mcg_hst}}
\end{figure}

The HEXTE measurement contains 92.6 ks of 800 cm$^2$ exposure on-source
and 80.6 ks off-source. Fitting with a single power law yields a best
fit for the index of $\Gamma$=2.18$\pm$0.09 with 20$-$50 keV flux of
2.70$\pm$0.11 $\times$ 10$^{-11}$ ergs cm$^{-2}$s$^{-1}$ and is shown
in Fig. \ref{fig.mcg_hst}.  The $\chi^2_\nu$ was 1.14 for 221 degrees of
freedom (The data from the two clusters were combined before fitting).
The best fit index is intermediate between the lower energy index
($\sim$1.7) and the higher energy value (2.7$\pm$0.4) mentioned above.
This is indicative of a more complex spectrum with a break at higher
energies. Indeed, Grandi et al.\markcite{grandi} (1997) have suggested
that a reflection component in conjunction with a cutoff power law
($\Gamma$=1.73$\pm$0.06, E$_C$=266$^{+90}_{-68}$) may be a better
representation of the MCG 8-11-11 spectrum from analysis of concurrent
ASCA and OSSE data taken 4.5 months before the {\it RXTE} observation.
Using a cutoff power law, the HEXTE values are $\Gamma$=1.90$\pm$0.35,
E$_C$=120.8$\pm$120.8 with a reduced chi-square of 1.17. The HEXTE
results are consistant with the ASCA/OSSE values.

\section{Conclusions}

Since its launch at the end of 1995, the HEXTE has performed extremely
well, with the one exception of the increased deadtime needed for
recovery after large energy deposits in the crystals. The tight control
of systematics allows for measurements of weak sources to better than 1\% of background
and for absolute timing at the submillisecond level.  HEXTE
observations of the Seyfert galaxy MCG 8-11-11 show that {\it RXTE} can
extract the source spectrum of weak AGN to 100 keV or more in 200 ks
observations.  Observations of the Crab Nebula and Pulsar verified the
{\it RXTE} absolute timing to significantly less than a millisecond,
and detected the break in the total Crab spectrum near 60 keV.
These examples confirm that users can expect to extract details of spectral
components that will lead to a better understanding of the processes and
geometries of x-ray sources using {\it RXTE}.

\acknowledgments

The HEXTE team would like to acknowledge the guidance of Prof. Laurence
Peterson in the development of HEXTE, the technical advice of Dr. James
Matteson, and the support of the {\it RXTE} mission by Drs. Alan
Bunner, Louis Kaluzienski, and others at NASA Headquarters over the
decade and a half from proposals to launch. The HEXTE development
depended upon the expertise and dedication of an exceptional technical
and programmatic staff, including Edwin Stephan Jr., Robert Howe,
Shirley Roy, Fred Duttweiler, Phillipe Leblanc, Charles James, George
Huszar, Tom Gasaway, Richard Bentley, Mathew Marlowe, Shalom Halevy,
Trevor Lillie, Georgia Hoopes, Ginger Beriones, and Paul Yeatman. We
thank the Scripps Marine Machine shop and Jack Kuttner.

We acknowledge the excellent work developing the HEXTE flight
electronics and flight software of the Perkin Elmer Applied Science
Division (now a division of Orbital Sciences Corporation) under the
management of Robert Kobiyashi and Robert Hertel.

We also would like to thank the Goddard Space Flight Center spacecraft
engineers for providing us with a superb, flexible, and intelligent
spacecraft that has met and exceeded all our expectations.  We
acknowledge the hard work and determination of the XTE Project office
through the many evolutions of the spacecraft and launch vehicle
concepts. Dale Schulz and the rest of the project team are to be
congratulated. We thank our colleagues on the instrument teams at GSFC
and MIT, and the XTE Science Working Group for excellent discussions of
technical and programmatic problems over the years.  Finally, we thank
the late Leon Herried for the software concept to operate the {\it
RXTE} in a flexible and straightforward manner. We also thank the
anonymous referee for suggesting improvements to the paper.  This work
was supported by NASA contract NAS5-30720.

\newpage

\begin{table}[ht]
\caption{HEXTE Instrument Characteristics}
\begin{tabular}{ll}
\tableline
\tableline
{\bf Characteristic} & {\bf Value}\\
\tableline
Energy Range &15 to 250 keV\\
Energy Resolution &$<$17\% at 60 keV\\
Time Resolution & 7.6 $\mu$s\\
Field of View &1$^{\circ}$ FWHM Hexagonal\\
Detector Material &NaI(Tl), CsI(Na)\\
Net Detector Area &1600 cm$^2$\\
Detection Sensitivity in 10$^5$ seconds, 3$\sigma$ &10$^{-6}$ photons cm$^{-2}$s$^{-1}$keV$^{-1}$ at 100 keV\\
Gain Control Stability&0.3\%\\
Source/Background Dwell Time &16 to 128 s\\
Calibration Source &$^{241}$Am (17 and 60 keV)\\
Detector Operating Temperature &17$^{\circ}$ to 19$^{\circ}$ C\\
Weight &400 kg\\
Operational Power &45 watts\\
Average Telemetry Rate &5,000 Bits/second\\
\tableline
\end{tabular}
\end{table}

\newpage

\begin{table}
\caption{HEXTE Phoswich Energy Resolution}
\begin{tabular}{rcrc}
\tableline
\tableline
Phoswich & $\Delta$E/E at 60 keV&Phoswich & $\Delta$E/E at 60 keV\\
\tableline
Cluster A, P0& 0.1426& Cluster B, P0& 0.1523\\
P1& 0.1639 &P1& 0.1676\\
P2& 0.1570&P2& 0.1654\\
P3& 0.1404&P3& 0.1458\\
\tableline
\end{tabular}
\end{table}

\begin{table}
\caption{$\chi ^2$ Values For Fitting Blank Sky Rates}
\begin{tabular}{llllcl}
\tableline
\tableline
Energy Band & Cluster & Net Rate & $\chi ^2$ & $\chi ^2$/18 DOF & Probability\\
\tableline
15$-$30 keV & A &  0.023$\pm$0.081 c/s & 20.91 & 1.16 & 0.284\\
          & B & -0.106$\pm$0.083     & 21.06 & 1.17 & 0.286\\
31$-$60 keV & A &  0.034$\pm$0.108 c/s & 11.14 & 0.62 & 0.888\\
          & B &  0.052$\pm$0.107     & 25.64 & 1.42 & 0.108\\
61$-$100 keV & A & -0.041$\pm$0.106 c/s& 16.05 & 0.89 & 0.589\\
           & B &  0.082$\pm$0.108    & 14.82 & 0.82 & 0.674\\
101$-$240 keV & A & 0.153$\pm$0.197 c/s& 26.72 & 1.48 & 0.084\\
            & B & 0.052$\pm$0.133    & 18.47 & 1.03 & 0.425\\
\tableline
\end{tabular}
\end{table}

\begin{table}
\caption{Crab Pulsar Ephemeris}
\begin{tabular}{ll}
\tableline
\tableline
$f$ & 29.8861513101 Hz\\
$\dot{f}$ & $-$3.75733 $\times$ 10$^{-10}$ Hz$^2$\\
T$_0$ & MJD 50311.000000119\\
R.A. & 83$^{\circ}$.633121\\
Dec. & 22$^{\circ}$.014460\\
\tableline
\end{tabular}
\end{table}

\clearpage

\newpage

\newpage

\newpage

\newpage

\newpage

\newpage

\newpage

\newpage

\newpage

\newpage

\newpage

\newpage

\end{document}